\documentclass[useAMS,usenatbib]{mn2e}
\usepackage{graphicx}

\title[]{}
\author[]{}

\title[Decaying binaries in dSph galaxies] {The tightening of wide binaries in dSph galaxies through dynamical friction 
as a test of the Dark Matter hypothesis} 
\author[X. Hernandez \& W. H. Lee] {X. Hernandez and William H.~ Lee \\ Instituto de Astronom\'{\i}a,
  Universidad Nacional Aut\'{o}noma de M\'{e}xico, Apartado Postal
  70--264 C.P. 04510 M\'exico D.F. M\'exico. \\} \date{Released 2007
  Xxxxx XX}

\pagerange{\pageref{firstpage}--\pageref{lastpage}} \pubyear{2007}

\def\LaTeX{L\kern-.36em\raise.3ex\hbox{a}\kern-.15em
    T\kern-.1667em\lower.7ex\hbox{E}\kern-.125emX}

\begin{document}

\label{firstpage}

\maketitle

\begin{abstract}
We estimate the timescales for orbital decay of wide binaries embedded
within dark matter halos, due to dynamical friction against the dark
matter particles. We derive analytical scalings for this decay and
calibrate and test them through the extensive use of N-body
simulations, which accurately confirm the predicted temporal
evolution. For density and velocity dispersion parameters as inferred
for the dark matter halos of local dSph galaxies, we show that the
decay timescales become shorter than the ages of the dSph stellar
populations for binary stars composed of 1 $M_{\odot}$ stars, for
initial separations larger than 0.1 pc. Such wide binaries are
conspicuous and have been well measured in the solar neighborhood. The
prediction of the dark matter hypothesis is that they should now be
absent from stellar populations embedded within low velocity
dispersion, high density dark mater halos, as currently inferred for
the local dSph galaxies, having since evolved into tighter binaries. 
Relevant empirical determinations of this
will become technically feasible in the near future, and could provide
evidence to discriminate between dark matter particle halos or
modified gravitational theories, to account for the high dispersion
velocities measured for stars in local dSph galaxies.
\end{abstract}


\section{Introduction} \label{intro}
Over the past few years the dominant explanation for the large mass to
light ratios inferred for galactic and meta galactic systems, that
these are embedded within massive dark matter halos, has begun to be
increasingly challenged. The lack of any direct detection of the
elusive dark matter particles, in spite of extensive and dedicated
searches, has led some to interpret the velocity dispersion
measurements of stars in the local dSph galaxies, the extended and
flat rotation curves of spiral galaxies (Milgrom \& Sanders 2003,
Sanders \& Noordermeer 2007, Nipoti et al. 2007, Famaey et al. 2007,
Gentile et al. 2007, Tiret et al. 2007, Sanchez-Salcedo et al. 2008),
the large dispersion velocities of galaxies in clusters, the
gravitational lensing due to massive clusters of galaxies, and even
the cosmologically inferred matter content for the universe, not as
indirect evidence for the existence of a dominant dark matter
component, but as direct evidence for the failure of the current
Newtonian and General Relativistic theories of gravity, in the large
scale or low acceleration regimes relevant for the above. Numerous
alternative theories of gravity have recently appeared, (TeVeS of Bekenstein
2004, and variations, Sanders 2005, Bruneton \& Esposito-Farese 2007, Zhao 2007, 
F(R) theories e.g. Sobouti 2007) now
mostly grounded on geometrical extensions of General Relativity, and
leading in the Newtonian limit to laws of gravity which in the large
scale or low acceleration regime, mimic the MOdified Newtonian
Dynamics (MOND) fitting formulas.

It appears that no matter what dark matter potential is inferred in
Newtonian dynamics, from observed rotation velocities or velocity
dispersion measurements, it could in principle be accounted for by a
suitably tuned alternative theory of gravity, resulting in an enhanced
gravitational relevance for the normal baryonic matter content of a
galactic system or collection of systems. However, even if the fit to
the resulting orbits where equally good under both hypotheses, they
reflect distinctly conflicting intrinsic physical realities, only one
of which can be true; either the space in question is teeming with
unseen exotic particles, or it is not.

Several recent studies have attempted to provide tests which might
decide between the dark matter hypothesis and modified gravitational
theories, focusing mostly on MOND vs. dark matter comparisons, not in
terms of resulting orbits, but by considering other higher order
gravitational effects. The derivative of the gravitational force leads
to tidal forces, with their role in limiting the sizes of satellites
and, in MOND, establishing escape velocities from satellites or
galaxies subject to an external acceleration field. Sanchez-Salcedo \&
Hernandez (2007) calculated tidal radii and escape velocities for
local dSph galaxies under MOND and dark matter, and comparing to
relevant observations, concluded a somewhat better fit under dark
matter. Wu et al. (2008) calculated the escape
velocity for the Milky Way under both MOND and dark matter, and
concluded that under both hypotheses the LMC appears as a bound
object, in spite of the recently high proper motion determined for
this object, given current observational
uncertainties. Sanchez-Salcedo et al. (2008) looked at the thickness
of the extended HI disk of the Milky Way from both angles, and
found a somewhat better fit to observations under MOND. Recently,
Skordis et al. (2006) studied the cosmic microwave background, Halle
et al. (2007) looked at the problem of the cosmic growth of structure,
Zhao et al. (2006) studied gravitational lensing of galaxies and Angus
et al. (2007) and Milgrom \& Sanders (2007) studied dynamics of
clusters of galaxies, all under MOND, all finding its description
of the problem as a plausible option, within the observational errors
of the relevant determinations. Most of the exploration in terms of alternatives to dark matter
has traditionally focused on MOND, but recently it has been found
at times to fail. For example, Zhao et al. (2006) find that when comparing across physical scales,
it either requires the inclusion of dark matter (as also found in Sanchez-Salcedo \& Hernandez 2007),
or the structural parameters of the theory need to vary from one galaxy to another. The 
problem is more general than a MOND vs. dark matter comparison, but actually encompasses 
the need to decide between dark matter and any alternative theory of gravity which attempts to
explain astrophysical observations in the absence of dark matter, e.g. MOND, or TeVeS or F(R)
type theories.

Here we explore a physical mechanism which might prove to be of help
in the present dark matter vs. modified gravity debate. For massive
bodies orbiting within dark matter halos, the individual two-body
interactions between the massive body and each dark matter particle
result in a net frictional drag which opposes the motion of the
massive body.  This is the well known dynamical friction effect. The
resulting timescales for dynamical friction in galactic systems are
typically in the Gyr range, and no direct observation of orbital decay
exists. However, the problem has been studied extensively, and it is
generally accepted that dynamical friction will result, for example,
in the eventual in-spiraling of the Magellanic clouds onto the Milky
Way. Dynamical friction is often considered as one of the main
mechanisms responsible for the hierarchical merger of galaxies in many
current standard cosmological structure formation models.

Studies of dynamical friction in dSph galaxies have been used to
constrain the density profiles of dark matter halos, for example, when
considering the decay timescales of orbits of globular
clusters. Through the explicit dependence of the problem on the
distribution function of dark matter particles, it has been shown that
orbital decay timescales longer than the lifetimes of the system can
be obtained for dark halo profiles characterized by constant density
cores rather than divergent cusps, e.g. Hernandez \& Gilmore (1998),
Read et al. (2006). Indeed, even though the orbits resulting from a
given baryonic mass distribution under a modified gravity theory and
those resulting from the dark matter hypothesis might be equally good
fits to observations, the dynamical friction problem will be
distinctly different under both scenarios. This has been realized, and
recently several authors (Sanchez-Salcedo et al. 2006, Nipoti et
al. 2008) have calculated dynamical friction decay timescales for
globular clusters in dSph galaxies under the dark matter hypothesis,
and under the MOND paradigm, where the stellar population itself acts
as a dragging background on the globular clusters. The conclusion is
that in the latter, dynamical friction timescales are shorter, perhaps uncomfortably shorter than the
system lifetimes, something that is easily avoided in the dark matter
scenario if the dark halos have central constant density cores.

In going to the stellar regime, the two components of a binary stellar
system, if embedded within a dark matter halo, will experience
dynamical friction. This will result in the progressive tightening of
the orbit, leading to the eventual merger of the two stars in the
absence of other effects. As is the case in the standard dynamical
friction problem of the orbital decay of a massive body, it is the
ratio of the orbital velocity to the velocity dispersion of the dark
matter particles that largely determines the strength of the
frictional drag and therefore the speed of the decay. One finds that
as the velocity dispersion of the particles responsible for the
dynamical friction drag becomes much larger than the velocity of the
body undergoing this effect, the drag rapidly tends to zero. Also,
this frictional drag is found to scale with the local density of dark
matter particles. For the large velocity dispersion values estimated
for the Milky Way halo, of order $220 / 3^{1/2}$ km/s, the effect on
binary stars is negligible. However, it is plausible that in going to
the highest density and lowest velocity dispersion dark matter halos
inferred, those of the local dSph galaxies, dynamical friction on
stellar scales might become relevant. Under a modified theory of
gravity without dark matter, there would be no particulate background
dragging down the binary orbit, and hence no orbital tightening for
binary stars.

The physical problem we will be treating is very similar to the 
study of the orbital evolution of a supermassive black hole binary due to interactions with a background
stellar population. The above has been studied extensively, for example in the paper by Quinlan (1996), with more recent in the works of Merritt (2000),
the comprehensive review of Merritt \& Milosavljevic (2005) and references therein, and that of Sesana et al. (2007).
In both problems we have a gravitationally bound binary with components of approximately 
equal mass, $M_{1}\simeq M_{2}$, embedded within a distribution of much lighter particles, $m \ll M_{1,2}$. The difference lies in the fact that
the supermassive black hole binary is always modeled as residing at the center of a galaxy, while the binary
stars studied here have a certain orbit within their host galaxy. The above results in the black
hole binary eventually depleting somewhat the central region of stars, and further
orbital evolution being determined (in the absence of gas, a key ingredient 
in the binary black hole case, which is absent from dSph local galaxies) by the rate at which
external stars can replenish the central regions. For the solar mass binaries moving slowly 
through their hypothetical dark matter halos as considered in this paper,
the underlying assumption of always finding them within a constant density background  of dark matter particles is valid.

Here we explicitly calculate the dynamical friction problem
for binaries within dark matter halos. We find that indeed, for the
density and velocity dispersion values inferred for the dark matter
halos of the local dSph galaxies, wide binaries would decay in
timescales shorter than the typical lifetimes calculated for the old
stellar populations of these systems of order 10 Gyr (e.g. Hernandez et
al. (2000), Lanfranchi et al. (2006)).

The prediction of the dark matter hypothesis for the distribution of
binary stellar separations in the local dSph galaxies is hence of a
much reduced upper limit separation, with respect to the Solar
neighborhood sample. In a modified gravitation scenario on the other
hand, abundant wide binaries could well be found in the local dSph
systems. Relevant binary population studies are currently technically
unfeasible, but will become possible in the near future, for example
through the GAIA satellite (Perryman et al. 2001).

The extreme low stellar densities make the local dSph galaxies ideal
probes for the effect being presented, as binary evolution due to
stellar encounters can be estimated to be a second order effect. The
total rate of collisions between two types of objects can be computed
as (see e.g., Binney \& Tremaine 1987):

$$
\nu_{\rm col}=f_1 f_2\left( \frac{n_{\rm c}}{ 10^{-4}{\rm pc}^{-3}}\right)^{2} \left( \frac{r_{\rm c}}{1 {\rm kpc}}\right)^{3} 
\left( \frac{\sigma}{5 {\rm km~s}^{-1}} \right)^{-1} \left( \frac{r_{\rm col}}{0.01 pc}\right) {\rm Gyr}^{-1},
$$ where $f_i \leq 1$ are the fractional number of systems of type $i$
in the core, $n_{\rm c}$ is the number density of such objects,
$r_{\rm c}$ is the core radius (within which the density is constant),
$\sigma$ is the velocity dispersion and $r_{\rm col}$ the distance of
closest approach (essentially the impact parameter, here taken to be
comparable to the binary separation, $a$, for a disruptive event). For
the typical numbers applicable to dSph galaxies, we find that the
rates are low enough to be ignored in a first approximation (although
a full population synthesis studie would probably need to consider
this more carefully).

In \S~2 we present the analytical estimates of dynamical friction
on the individual stellar components of a binary system, which we
calibrate through extensive numerical simulations in \S~3.  In
\S~3 we also test the predicted scalings found in \S~2, and
validate the overall analytic approach, which allows us to present
general in-spiraling evolution formulas. A discussion of our results
in the context of the local dSph galaxies, together with our
conclusions are given in \S~4.

\section{Analytic calculations}\label{analytical}

The standard approach to the study of dynamical friction problems
usually starts from the Chandrasekhar formula. This considers the
specific case of a massive body moving in a straight line through an
infinite, constant density medium of small particles having a
Maxwellian distribution function. Although numerical studies have
generally validated its use in the case of circular orbits within
finite systems, in our case we deviate significantly from the
underlying assumptions, making it necessary to develop a detailed
treatment from even more basic principles. The problem is that
generally one considers a massive body orbiting on large circular
orbits, of radius comparable to that of the halo of particles. The
orbital period of the massive body is so long that by the time it
completes a revolution, it returns to a region where the distribution
function has long since returned to its undisturbed form, making the
hypothesis leading to the Chandrasekhar formula of a fixed density and
distribution function for the background medium valid.

Here we will think of a stellar binary in which the components revolve around each other in a circular orbit. This binary is embedded within an infinite medium of small dark matter particles (the large scale DM halo) which produce a constant background density, $\rho_{0}$, and obey a isotropic Maxwellian distribution
function $f_{0}(v) \propto \exp(-v^{2}/2 \sigma^{2})$ everywhere.  The
presence of the binary alters the distribution function and density
distribution, leading to the build up of a perturbation, which will
result in the frictional drag.  We shall treat the problem through the
first order perturbation of the distribution function.

To get a first idea of what the binary will do to the initially
constant density dark matter halo, we shall begin by looking at the
effect of a single star, at rest with respect to the background.  This
single star will contribute a perturbation to the potential felt by
the dark matter given by:

\begin{equation}
\Phi_{1}(r)=\frac{- G M}{r},
\end{equation}
where $M$ is the mass of the star and $r$ is a radial coordinate
centered on the star in question.  The problem evidently has spherical
symmetry. We shall now approximate the distribution function as:
$f(r,v)=f_{0}(v)+\epsilon f_{1}(r,v)$, to which there will correspond
a perturbed total density $\rho(r)=\rho_{0}+\epsilon \rho_{1}(r)$ due
to an overall potential, $\Phi(r)=\Phi_{0}(r)+\epsilon \Phi_{1}(r)$.
The full distribution function will satisfy the Boltzmann equation:

\begin{equation}
\frac{\partial f}{\partial t} + \vec{v} \cdot \vec{\bigtriangledown} f
- \vec{\bigtriangledown} \Phi \cdot \vec{\bigtriangledown_{\vec{v}}}
f=0.
\end{equation}

We will now look for a stationary solution and keep only terms to first
order in the perturbation. Taking $\Phi_{0}=0$, the familiar Jeans swindle, 
we obtain for the radial component:

\begin{equation}
v \frac{\partial f_{1}}{\partial r}= \frac{G M}{r^{2}} \frac{\partial
  f_{0}}{\partial v}.
\end{equation}

Using the explicit dependence of $f_{0}(v)$, we can write the right
hand side as of equation(3) as:

$$
\frac{-v}{\sigma^{2}}\frac{G M}{r^{2}} f_{0},
$$ after which we can integrate equation (3) over velocity space to
yield

\begin{equation}
\frac{d \rho_{1}}{d r}= \frac{-G M}{\sigma^{2} r^{2}} \rho_{0}.
\end{equation}

The above equation can be readily solved to yield the density
perturbation induced upon the dark matter distribution by the single
star at rest as:

\begin{equation}
\rho_{1}(r)=\frac{G M}{r \sigma^{2}} \rho_{0} \; \; \; = \left( \frac{v_{e}}{\sigma} \right)^{2} \rho_{0}, 
\end{equation}
where we have introduced an equivalent circular velocity due to the
star, $v_{e}^{2}=GM/r$. To first order, we can think of the response
of the dark matter halo to the presence of a wide binary, where the
orbital velocity is lower than $\sigma$, as being composed of two
density enhancements, centered upon each of the stars in the binary,
each described by equation (5).  Indeed, a close analogy can be found
when calculating the dynamical friction due to a gaseous medium.
Sanchez-Salcedo \& Brandenburg (1999) find that in the sub-sonic case,
when the perturber moves at speeds lower than the sound speed of the
medium, the response is in fact an approximately spherical
perturbation centered on the perturber.

Hence, the resulting density enhancement will produce no gravitational
force on the stars in the binary, being a spherical density
enhancement centered upon each of the stars. However, it is a mistake
to conclude that there will be no dynamical friction, because the
identity of the dark matter particles making up each density
enhancement is not fixed. Each dark matter particle only briefly forms
part of the density response of the dark matter halo. In the absence
of the binary, the angular momentum of the dark halo will be zero,
whilst in its presence a steady state ensues, where two density
enhancements, as given by equation (5), will follow it. As these two
enhancements have to be constantly reformed, we can calculate the
angular momentum loss for each star in the binary star as:

\begin{equation}
\dot{L}= \frac{M_{e} V_{o} R_{o}}{\tau}.
\end{equation} 

In the above equation $M_{e}$ is the mass of each one of the density
enhancements centered on the stars in the binary, $V_{o}$ and $R_{o}$
are the binary orbital velocity and orbital radius, and $\tau$ is a
characteristic timescale over which the density enhancement is being
replenished. To estimate $M_{e}$ we have to integrate equation (5),
out to a certain maximum radius, $R_{Max}$. This can not be larger
that $R_{o}$, as then the two density enhancements would overlap, and
the reaction of the dark halo to the presence of one of the stars in
the binary would be erroneously calculated as being still determined
by the distant component. $R_{Max}$ could be smaller than $R_{o}$, in
cases where inertial forces introduce a smaller truncation, this we
shall consider in the appendix. As a first approximation we will take
$R_{Max}$=$R_{o}$, to integrate equation (5), to obtain $M_{e}$ as:

\begin{equation}
M_{e}=2 \pi \frac{G M}{\sigma ^{2}} R_{0}^{2} \rho_{0} \; \; \; = \frac{3}{2} M \left( \frac{v_{h}}{\sigma} \right)^{2},
\end{equation}
where we have introduced an equivalent circular velocity due to the
dark halo at the binary scale, $v_{h}^{2}=G M_{h}/R_{o}$, with
$M_{h}=(4\pi/3) \rho_{0} R_{0}^{3}$, the dark halo mass within a
sphere of radius $R_{0}$ under unperturbed conditions. To estimate
$\tau$, we can think of the dissipation of the density enhancement to
be of order $R_{0}/\sigma$. However, in the absence of the binary, the
density enhancement will first stop turning, and only later
dissipate. In order to estimate the rate at which the binary loses
angular momentum, it is more exact to think of $\tau$ as given in
terms of the binary orbital period, $\tau = \alpha / \Omega$, with
$\Omega$ the orbital frequency of the binary and $\alpha$ a
dimensionless scaling factor.  With these hypothesis, we can obtain
the rate of loss of angular momentum for each component of the binary
from equation (6) as:

\begin{equation}
\dot{L}= \left( \frac{4 \pi}{\alpha}\right) \left( \frac{G M}{\sigma} \right)^{2} \rho_{0} R_{0},
\end{equation}
which we can write in terms of the angular momentum of each star in
the binary $L_{b}=(2 G M^{3} R_{0})^{1/2}$ and the particle crossing
time for the dark matter particles over the radius of the binary,
$\tau_{c}=R_{0}/\sigma$, as:

$$
\dot{L}=\left( \frac{3}{2^{3/2} \alpha} \right) \left( \frac{L_{b}}{\tau_{c}} \right)
\left( \frac{v_{h}^{2} v_{e}}{\sigma^{3}} \right).
$$

If we now make the assumption that the orbital decay of the binary
proceeds slowly in terms of the orbital period, describing a very
closed spiral, we can think of the orbit as circular throughout the
evolution. This allows to obtain the temporal evolution of the binary
radius by deriving the expression for $L_{b}$ with respect to time,
and equating it to equation (8),

$$
\left(G M^{3} \right)^{1/2} \frac{\dot{R_{o}}}{(2 R_{0})^{1/2}} =  
\left( \frac{4 \pi}{\alpha}\right) \left( \frac{G M}{\sigma} \right)^{2} \rho_{0} R_{0},
$$
giving:

\begin{equation}
\dot{R_{0}}=\left( \frac{2^{5/2} \pi }{\alpha} \right) 
\left( \frac{G^{3/2} M^{1/2} \rho_{0}}{\sigma^{2}}\right) R_{0}^{3/2},
\end{equation}
or

$$
\dot{R_{0}}= \frac{3}{\alpha} V_{0} \left( \frac{v_{h}}{\sigma} \right)^{2} \;\; = \frac{4 \pi}{\alpha}V_{0} 
\left( \frac{\tau_{c}}{\tau_{ff}} \right)^{2}.
$$

The last expression for $\dot{R_{0}}$ is given in terms of the
gravitational free fall time of the unperturbed background. Since
$\tau_{c}$ scales with the radius, and for the whole halo $\tau_{ff}
\sim \tau_{c}$, given that the radius of the binary is much smaller
than the size of the dark matter halo, we see that the rate of radial
decay will be much slower than the orbital velocity of the binary. In
the following section we shall see that $\alpha$ is of order
$10^{-3}$, and thus the in-spiraling will therefore precede along
tight spirals, as assumed previously. To order of magnitude, the
parenthesis in the last expression can also be replaced by the ratio
of the binary radius to the dark halo size. The solution to equation
(9) is trivial, and gives the temporal evolution of the binary radius
as:

$$ t= \left( \frac{\alpha}{2^{3/2} \pi} \right)
\frac{\sigma^{2}}{\rho_{0} M^{1/2} G^{3/2}}
\left(\frac{1}{R_{0}(t)}-\frac{1}{R_{0}(t=0)} \right)
$$

What remains now is to evaluate $\alpha$. This will be done in the
following section, where high resolution numerical simulations are
used to derive $\dot{R_{0}}$ for a few specific cases, from which we
calibrate $\alpha$. A more detailed treatment of the dark halo
response to the presence of the rotating binary, with full account of
the inertial forces present, is included in the appendix. The
calculations there shown, together with the numerical results of the
following section, will be seen to validate the assumptions going into
the derivation of equation (9).

\section{Numerical calculations}\label{numerics}

We have carried out a series of numerical experiments to directly
compute the rate at which the binary separation decays under the
influence of dynamical friction. 

The standard setup consists of a binary with identical components,
with mass $M_1=M_2= 1 M_{\odot}$, placed in a circular orbit with
semi-major axis $a=2 R_{0}$. The three-dimensional volume in the
vicinity of the binary, out to a radius $R > a$, is filled with dark
matter particles of mass $m \ll M_{1,2}$ which, in the absence of the binary,
produce a constant mass density $\rho_{\rm DM}=1 M_{\odot}$~pc$^{-3}$
and have an equilibrium Maxwellian distribution function with
isotropic velocity dispersion $\sigma$=5 km/s, in accordance with the assumptions in our analytic derivation in \S~\ref{analytical}. These parameters in
core radius and density are in
the typical ranges estimated for the central regions of the dark
matter halos of local dSph galaxies, where the bulk of the stellar
populations reside, as inferred through modeling of the observed
kinematics of their stars e.g. Koch et al. (2007). The corresponding typical values for 
the velocity dispersion in the better studied local dwarf spheroidals with the earliest discovery dates, are
of around $\sigma$=10 km/s. However, we will calculate the case for  $\sigma$=5 km/s,
rather corresponding to the more recently discovered of such systems, e.g. Gilmore et
al. (2007) and references therein. This is done to set ourselves in the
astrophysical case where equation (9) predicts the highest decay rates.

During the simulation, each star feels the gravitational effect of the
companion and of all the dark matter particles directly (no
approximations, other than a gravitational softening length to avoid
excessive accelerations in the case of close encounters, are
used). The dark matter particles feel only the presence of the two
stars (i.e., they are not self-gravitating), allowing for a large
number of particles to be used, since the computation time scales with
the number of dark matter particles, $N_{\rm DM}=4 \pi R^{3} \rho_{\rm
  DM}/3 m$. This is substantially different in terms of computational
requirements than, for example, the interaction of galaxies (Hernandez
\& Lee 2004) or individual stars (Lee \& Ramirez-Ruiz 2007), in which
the self gravity of the various components and gas dynamics are
important ingredients one needs to consider when doing detailed
modeling.

The numerical integration is performed with a Runge-Kutta algorithm
accurate to fourth order, and we have checked that in the limit of
vanishing dark matter mass density, the binary remains stable, with
the separation remaining constant.

The distribution of orbital separations in local
binaries has been well studied, e.g. Griffin (2006), with detailed modeling
yielding orbital separation distributions well fitted by a uniform distribution in
log(a) for a in the interval 10Au to 0.5 pc, e.g. Eggleton et al. (1989).
We run four cases, uniformly spaced in log ($R_{0}$), at initial values
of a=$(10^{-4}, 10^{-3}, 10^{-2}, 10^{-1})$ pc. The upper value was chosen to be consistent
with high values found for local samples, in fact, we remain a factor of 5 below the upper range
of the distribution of Eggleton et al. (1989). It is not known what the intrinsic binary separations in
the dSphs might be, but we take it as plausible that the local sample might serve as an initial
zero order approximation. 

In the absence of the binary, the dark matter placed inside the sphere
of radius $R$ will leak out of it isotropically (since the initial
spatial distribution is homogeneous, and the velocities are isotropic)
at a rate given by

\begin{equation}
\dot{M}_{\rm leak}= (8 \pi)^{1/2} \, \rho_{\rm DM} \, \sigma \, R^{2},
\label{eq:leakrate}
\end{equation}
so that their number would drop to zero if nothing were done. In fact we
remove particles when they cross the outer boundary at $R$, and inject
new particles over the surface of the sphere at twice the rate given
in equation~(\ref{eq:leakrate}). Half of them are removed essentially
immediately because their velocity vector points outward, while the
remaining move in. We have verified that the expression in
equation~(\ref{eq:leakrate}) is correct numerically by computing a test
case in which the stars are absent. The number (and mass) density of
the dark matter particles remains constant with the algorithm
described above to within sampling errors given by discretization
effects. 

Ideally one would hope to have a sphere around the binary that is much
larger than the separation between the stars so as to allow for the
correct trajectories of the particles that are injected. The
computational cost at a fixed particle number density scales as
$R^{3}$ and so we have chosen a trade-off, with $R=5a$.  
This choice sensitive to the velocity dispersion, $\sigma$. At the start of the calculation there will be an initial transient because the binary
suddenly feels the effect of a homogeneous distribution of matter
around it. After a crossing time, $\tau_{c}$, the
majority of the initial particles will have left the sphere, and it
will be filled with ones that have been injected at its boundary. For
the trajectory of an infalling particle to be computed realistically,
the escape velocity at radius $R$ must be small compared with the
typical velocity (the dispersion), i.e., we must have:
\begin{equation}
v_{\rm esc}(R)= \left(\frac{4GM}{R}\right) ^{1/2} \leq  \sigma,
\label{eq:sigmaesc}
\end{equation}
where $2M$ is the total mass in the stellar binary. For our
parameters, with $\sigma=5$~km~s$^{-1}$, and $R=5a$ we find that
equation~(\ref{eq:sigmaesc}) is satisfied for $a \geq 10^{-2}$~pc. At
smaller radii not taking the infall trajectories properly could modify
the interaction with the binary. However, as will be seen below, this
falls in the range of parameter space where the decay rate is already
very small, and thus does not alter our conclusions significantly. We did
nevertheless, for small binary separations ($a=10^{-4}, 10^{-3}$~pc)
perform simulations where $R$ was greater (up to $R=15a$) in order to
carefully check for numerical convergence of the actual decay rate.

A large number of dark matter particles is required in order to successfully
model the effect of dynamical friction on the binary and reduce the
level of noise due to discretization. We have computed the evolution
of the binary for fixed values of the initial separation, $a$, at
increasing levels of resolution, with $N_{\rm DM}$ typically ranging
from $2 \times 10^{3}$ to up to $10^{6}$. The number of particles
required for numerical convergence of the solution correlates with the
choice of $a$, and is $N\simeq 10^{6}$ for the largest binary
separations (0.1~pc) we considered.

\begin{figure}
\includegraphics[angle=-90,scale=0.5]{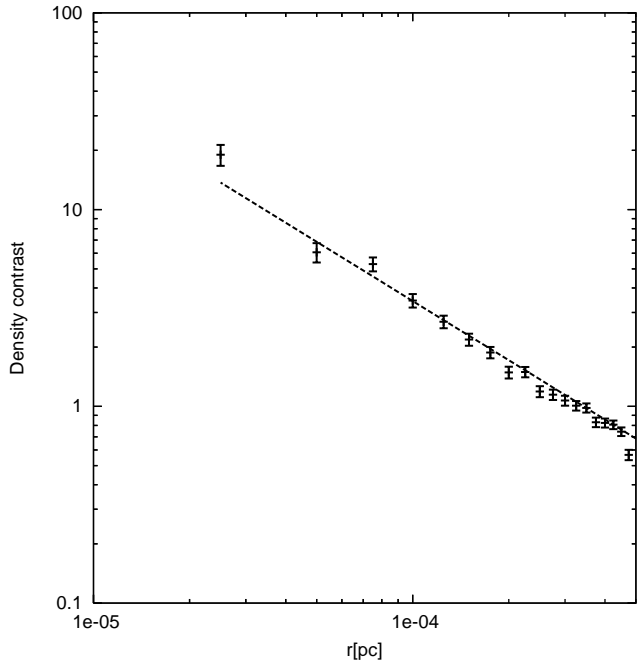}
\caption{Dark matter density profile around a single star as a
  function of distance from it, after a steady state has been
  achieved, with error bars showing sampling errors.  The straight
  line is the anaylitcal prediction of equation (5).}
\label{fig:singleprof}
\end{figure}

We find that after a relaxation time of order $\tau_{c}$, the simulation settles to a steady state
solution in which the decay rate is constant. When seen in a reference frame rotating with the binary,
this steady state is characterized by two static density enhancements,
very closely centered on each of the stars. Figure~(\ref{fig:singleprof}) then
shows the density enhancement as a function of distance to a single
star, where we have averaged over spheres centered on the position of
each star, in the rotating frame of the binary. The straight line
gives the prediction of equation (5), strictly, derived for a static
star.  We see that out to the radius of the binary, $R_{0}= 0.0005 pc$
in this case, the actual response of the dark halo does not deviate
from the analytic prediction beyond sampling errors, which it matches
in both the amplitude and the radial scaling. We have also
checked that the form and amplitude of these density enhancements are
in fact constant with time.

When calculating the decay rate over long periods of time, at very low 
resolutions (i.e., number density of particles) the result is not
converged, showing fluctuations in the decay and sometimes even
reversals. As the number of particles in the computational volume
increases the decay is smoother and becomes monotonic. The correct solution when converged is in fact attained from below in
the decay rate, making our estimates conservative absolute lower
bounds.

We have hence found through direct N-body simulation that the
expectations of section (2) are verified, in the regime where $V_{0}
\leq \sigma$. The response of the dark halo is in accordance with the
assumptions leading to the derivation of equation (9). It would
therefore appear to follow that the mechanism of angular momentum loss
for the binary has been correctly identified, and that the numerically
calculated decay rates should also follow the scalings of equation
(9). A careful computation of a few particular cases was then used to
calibrate the $\alpha$ parameter appearing in equation (9).

The results of section (2), and in particular equation (9), indicate
that the binary decay rate should scale linearly with the background
matter density and with the square root of the total binary mass. We
have tested both of these predictions by varying $\rho_{\rm DM}$ by a
factor of 3 to both larger and smaller values, at a fixed particle
number density, and the total mass $2M$ by a factor of 4 (also to
higher and lower values) and find that this is indeed the case, with
$\dot{a} \propto \rho_{\rm DM} M^{1/2}$ (in all cases we maintained a
binary mass ratio of unity).

\begin{figure}
\includegraphics[angle=0,scale=0.43]{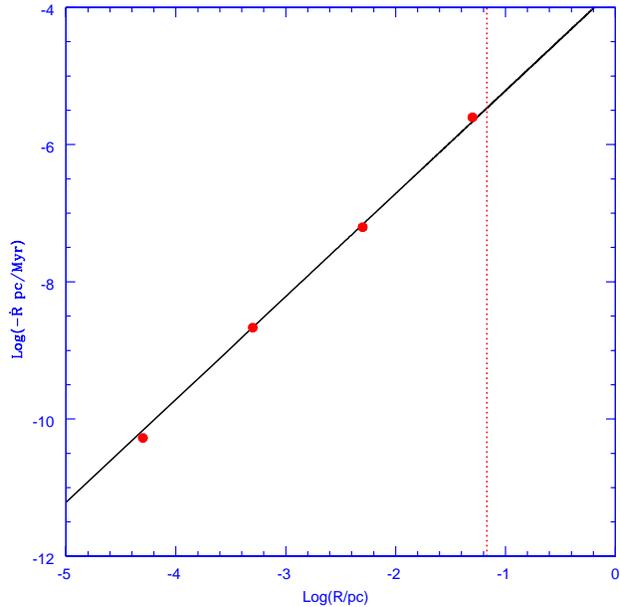}
\caption{The dots show four numerically calculated decay rates, for a
  halo density of $1 M_{\odot} pc^{-3}$ and a dark matter velocity
  dispersion of $5 km/s$. The solid line gives the analytical
  prediction of equation (9), for $\alpha=1.07 \times 10^{-3}$. The
  vertical dotted curve gives the binary radius beyond which $R_{0}/(2
  \dot{R_{0}})$ becomes shorter than 10 Gyr.}
\label{fig:Drate}
\end{figure}

In figure (\ref{fig:Drate}) we show with filled circles the best
estimate of the decay rate for four values of the initial separation
of the binary.  We have spanned four orders of magnitude in initial
separation, and run each simulation long enough to clearly establish
the decay rate, although the actual value of $R_{0}$ decreased only by
a small factor. The solid line gives the prediction of equation (9), a
decay rate which scales with $R_{0}^{3/2}$, where we have calibrated
$\alpha=1.07 \times 10^{-3}$ to obtain a best fit to the numerically
estimated decay rates. The excellent agreement of the simplified
analytical calculation, over four orders of magnitude against the
simulations is encouraging of the physical scenario presented in
section (2) as an explanation to the dynamical friction of binaries,
in the range of parameters relevant to dSph galaxies. This highlights
the inaccuracies incurred by assuming that only the direct
gravitational force of the induced wake determines the dynamical
friction problem; the sustaining of the wake itself, even if
spherically symmetrical about the perturber, also results in a drag of
linear or angular momentum.  This becomes particularly obvious in the
sub-sonic regime ($V< \sigma$), where the wake becomes spherical and
the direct gravitational drag vanishes.

It is interesting to compare equation (9) and the findings of this section with results for 
the orbital tightening of binary black holes. Recently, such studies
include the effects of gas, and of the depletion of stars from the central regions,
where the black hole binary resides e.g. Merritt \& Milosavljevic (2005). However in the first such studies, 
e.g. the thorough numerical work of Quinlan (1996), the tightening of a black hole binary was calculated 
including only the presence of a constant background of stars, essentially the situation we model here.
It is reassuring that these could be highly accurately fitted by
a function which satisfies equation (9), plus the inclusion of a further
factor of $(V_{0}/ \sigma) ln (\sigma^{2}/V_{0}^{2})$. It is possible that we have missed this factor,
because over the range of orbital separations we modeled numerically it varies only by a 
factor of a few. The analytical developments of section (2) provide a physical understanding
to the dominant scalings of the functional fits to the numerically calculated black hole binary decay rates 
of e.g. Quinlan (1996), who considered a wider range of parameters. The validity of equation (9) is restricted
to a certain range in $V_{0}/\sigma$ (see the appendix), over which there is excellent agreement with the numerical simulation performed here, and to those of black hole binary numerical simulations over the
range of parameters where they can be compared. The missing factor noted above becomes relevant only in the range where
$V_{0}/\sigma$ becomes larger. It is in this 'supersonic' regime where the simplifying assumptions behind equation (9) 
break down, most importantly ignoring the truncation of the density enhancements due to inertial forces, and the 
imposing of a stationary solution. These effects are hard to consider
analytically, but from comparison to the black hole
experiments, should be credited as the origin of the additional factor  $(V_{0}/ \sigma) ln (\sigma^{2}/V_{0}^{2})$ in Quinlan (1996).

The vertical dotted line in this last figure gives the initial value
of $R_{0}$ beyond which the typical timescale over which the orbit is
substantially modified, $\tau_{dec}= R_{0}/(2 \dot{R_{0}})$, becomes
larger than 10 Gyr, $R_{0}=0.067$ pc. The value of 10 Gyr was chosen
to reflect a typical value for the ages of the old component (in some
cases the only component) of the stellar populations in local dSph
galaxies, as inferred through the direct statistical modeling of
their resolved HR diagrams, e.g. Hernandez et al. (2000), Lanfranchi et
al. (2006). We therefore see that for wide binaries, the decay
timescales in dSph galaxies can become shorter than the lifetimes of
these systems. Although in the low density and high dispersion
velocity conditions inferred for the dark matter halo in the solar
neighbourhood the effects of dynamical friction on any stellar binary
are absolutely negligible, we have shown that the effect should become
relevant in local dSph.

\section{Discussion and Conclusions}\label{ccl}

The direct study of the distribution of binary separations in even the
closest local dSph currently remains beyond the scope of technical
feasibility, even with the HST, but only very slightly so (G. Gilmore,
private communication). It is certain that with the next generation of
space telescopes, the undertaking of detailed determinations of binary
separation distributions in local dSph galaxies will become possible.

The interpretation of whatever such studies reveal might be ambiguous.
If no binaries with separations larger than the limits we derived in
this paper are found, from the point of view of the dark matter
hypothesis, the results will be seen as a new and independent
validation of the physical reality of dark matter. From the point of
view of modified theories of gravity, the absence of wide binaries
will be interpreted as a reflection of initial conditions and a
different intrinsic mechanism for binary star formation, to what gives
rise to the solar neighborhood or local Galactic halo population,
perhaps unlikely given the universality of a kindred distribution, the
stellar IMF. If on the other hand plentiful wide binaries were to be found
in local dSph galaxies, the dark matter scenario would be very
seriously challenged. Also, given the linear dependence of the decay rates on the
local dark matter densities of equation (9), a strong galactic radial gradient in
the maximum binary separations would be expected in any hypothetical strongly cusped
dark matter halo, of the type suggested by recent cosmological computer simulations, e.g. 
Navarro et al. (1997).

The possibility of appreciably altering an initial distribution of
binary separations, of forming new wide binaries through stellar
capture, or of continuous replenishing through the partial disruption
of tighter binaries, as it happens in globular clusters, is in this
case not available. Although the velocity dispersions of stars are
comparable to what one finds in globular clusters, in dSph galaxies
these do not reflect the potential produced by the stars themselves,
but the overwhelming dynamical dominance of the dark matter (under such a hypothesis). The stars
therefore move as fast as they do in globular clusters, but the volume
densities are very substantially lower. The number of stars in a
typical dSph is of order $10^{6-7}$, about 10 times larger than in
globular clusters, the volume occupied by the stellar population in a
dSph however, is of order $(2 {\rm kpc})^{3} $, some million times
greater than the case in globular clusters, of order $(20 {\rm
  pc})^{3}$. This makes the usual mechanisms operating in globular
clusters completely inefficient. Also, for the test to be meaningful, one requires the 
use of main sequence stars, as strong mass loss over the red giant phase and beyond would result 
in the orbital widening of the binary, e.g. Valls-Gabaud (1988).

We have also assumed the binary is at rest. In reality, it will find
itself on an orbit within the dSph galaxy, and will therefore see the
dark matter particles as approaching with a certain velocity, which
will change in direction and magnitude over the course of the
orbit. Observational determinations of the orbits of stars in dSph
galaxies generally concur in assigning relatively elliptical orbits to
these stars. The typical ratios for the maximum to minimum radii for
the orbits being 0.3. Over such elliptical orbits, the binary will
spend the greater fraction of its orbital period moving very slowly
near its apocenter. For a small fraction of the time, it will be
dashing past its pericenter. This means that the calculations we have
performed will be valid over most of the galactic orbital period of
the binary. During the short lived, fast moving phase, the orbital
decay due to dynamical friction could be reduced, as the outer regions
of the density enhancements will be 'blown away' to some extent. In
any case, even during this fast moving phase, the galactic orbital
speeds will still be close to $\sigma$, so it is safe to assume the
effects of considering the galactic orbit for the binary will be minor
on the final dynamical friction timescales we have estimated
here. More important than the above effect, and making our estimates
of decay rates fall somewhat below the true values, is the lack of
self gravity for the dark matter particles in the simulations from
which we calibrated $\alpha$. If this effect were included, the
density enhancements about each star would become more massive and
coherent, resulting in increased dynamical friction effects.

We have presented a new and independent test of the dark matter
hypothesis. The empirical application is currently unavailable, but
will become so in the near future. This gives even grater relevance to
the continual development of increasingly advanced studies of the
stellar populations of the local dSph galaxies, they might hold the
clue to the dark matter mystery in more senses than formerly
appreciated.

\section*{acknowledgments}

The authors would like to thank the constructive criticism of an 
anonymous referee in helping 
to reach a more complete and clearer version of this paper.
This work was supported in part through CONACyT (45845E),
and DGAPA-UNAM (PAPIIT IN-113007-3 and IN-114107).

\section*{Appendix}

In this appendix we approximate for the density enhancement around
each of the two rotating stars in a binary at rest within a dark
matter halo, strengthening the physical understanding of the excellent
agreement between the analytic expectations of \S~2 and the
numerical experiments of \S~3.  The notation is as in \S~2, 
we shall proceed by writing the third Jeans equation in a
reference frame which rotates with the binary with angular frequency
$\Omega$, linearizing with respect to the density enhancement, and
looking for a stationary solution (in the rotating frame). The third
Jeans equation in the rotating frame, under cylindrical coordinates,
$(R, \theta, z)$, will read:

$$
\rho \frac{\partial \overline{v_{j}}}{\partial t} + \rho \overline{v_{i}} \frac{\partial \overline{v_{j}} }{\partial x_{i}}=
-\rho \frac{\partial \Phi_{ef}}{\partial x_{j}}- 2\rho \overline{(\vec{\Omega} \times \vec{v} )_{j}} - 
\frac{\partial(\rho \sigma^{2}_{ij})}{\partial x_{i}} \;\;\;\;\;\;\; (A1).
$$

In the above equation $\overline{X}$ refers to the mean (or expected)
value of $X$, the second term in the RHS gives the Coriolis force, and
the centrifugal force is included in $\Phi_{ef}=\epsilon
\Phi_{1}-|\vec{\Omega} \times \vec{r}|^{2}/2$, where we have again
taken the undisturbed potential to be zero, and $\Phi_{1}$ is the
potential due to each star.

We now take $\rho=\rho_{0}+\epsilon \rho_{1}$, $\overline{v_{i}}
=\overline{v_{0i}}+\epsilon \overline{v_{1i}}$ for all $i$.  Imposing
a stationary solution with no flows beyond $\overline{v_{0}}=(0,
\Omega R, 0)$, we get for $j=\theta$:

$$
0=-\frac{\rho_{0}}{R} \frac{\partial \Phi_{1}} {\partial \theta} -\sigma^{2}\left( \frac{\partial \rho_{1}}{\partial R} 
+ \frac{\partial \rho_{1}}{R \partial \theta} +  \frac{\partial \rho_{1}}{\partial z}  \right) \;\;\;\;\;\;\; (A2)
$$

Limiting the analysis to any plane $z=cst$, along any line $R=cst$,
we get for an isotropic velocity for the dark matter particles,

$$
 \frac{\partial \rho_{1}}{\partial \theta}=- \frac{\rho_{0}}{\sigma^{2}} \frac{\partial \Phi_{1}} {\partial \theta} \;\;\;\;\;\;\; (A3).
$$

Equation ($A3$) is analogous to equation (4), and shows that the
density enhancement on each star, even under the full rotation
description, will follow the angular variations in $\Phi_{1}$, i.e. it
will remain centered on the stars. This is a result of having assumed a
static solution with no flows, valid for orbital velocities not
substantially exceeding $\sigma$, and seen to hold through the
numerical simulations of \S~3.

Under the same assumptions, for $j=R$ equation $(A1)$ evaluated at
constant height and angular direction yields:

$$
0=- \rho_{0} \frac{GM}{(R-R_{0})^{2}} + \frac{\rho_{1} \Omega^{2} R}{2} -2\rho_{1} \Omega^{2}R -\sigma^{2} 
\frac{\partial \rho_{1}}{\partial R}    \;\;\;\;\;\;\; (A4)
$$

Writing ${\cal R}= R-R_{0}$ we can write the previous equation as:

$$
\frac{\partial \rho_{1}}{ \partial {\cal R}} = -\rho_{0} \frac{G M}{\sigma^{2} {\cal R}^{2}}
- \rho_{1} \frac{3}{2} \left( \frac{\Omega}{\sigma}  \right)^{2} ({\cal R} +R_{0}) \;\;\;\;\;\;\; (A5).
$$

We see that we have again obtained equation (4) the density
enhancement centered on each star of \S~2, with the addition of
an extra term which scales with the distance to the center of the
coordinate system, the center of the binary, and which is proportional
to $(\Omega / \sigma)^{2}$. This last term effectively contributes
with a radial exponential cut off to the density enhancement
calculated in \S~2, due to the inertial forces of the spinning
binary.

For small values of $R$ it is the first term on the RHS of equation
($A5$) which dominates, and we find the density enhancement of \S~2. 
For large values of $R$ the second term dominates, imposing an
effective cut off radius at a distance $R_{cut} \simeq \sigma /
\Omega$. For the approach of \S~2 to be valid, we need
approximately $R_{cut} \geq R_{0}$. With $\Omega= V_{0}/R_{0}$, this
condition reduces to:

$$
\frac{\sigma}{V_{0}} \geq 1,
$$
for the regime over which inertial forces do not dominate to the point
of truncating the density enhancements within a typical distance
$R_{0}$. Also, we need the dispersion timescales of the density
enhancements due to particle streaming not to dominate over the
turning timescale of binary orbit, i.e., approximately

$$
\frac{\alpha R_{0}}{V_{0}} \leq \frac{R_{0}}{2 \sigma}.
$$

Combining these last two conditions we obtain the range of validity for
the analytic decay rates of \S~2 as:

$$
\frac{1}{2 \alpha} \geq \frac{\sigma}{V_{0}} \geq 1  \;\;\;\;\;\;\; (A6).   
$$

For the values of $M$, $R_{0}$ and $\sigma$ used in \S~3, we
see that equation $(A6)$ is satisfied (recall the fit to $\alpha$ =
1.07 $\times 10^{-3}$) for all the numerical experiments, which
hence, yielded results in good agreement with the expectations of
\S~2.

\end{document}